\documentclass[preprint,12pt,authoryear]{elsarticle}

\usepackage{amssymb}
\usepackage{amsmath}

\usepackage{lineno}

\journal{Earth and Planetary Science Letters}

\begin{document}

\begin{frontmatter}



\title{Fluid flow and hydration in oceanic lithosphere: Insights from theory and numerical investigation} 


\author[a]{Amy Ferrick} 
\author[a]{Jun Korenaga}

\affiliation[a]{organization={Department of Earth and Planetary Sciences, Yale University},
            city={New Haven},
            postcode={06511}, 
            state={CT},
            country={USA}}

\begin{abstract}
The oceanic mantle lithosphere has considerable potential to store chemically bound water, thereby being an important factor for the deep water cycle. However, the actual extent of hydrous alteration in such mantle rocks is debated. Geodynamic modeling has the potential to directly predict the extent of fluid flow through oceanic lithosphere, and, in turn, the extent of serpentinization. By comparing theory and numerical simulations, we demonstrate that conventional geodynamic models are inherently inconsistent with the physics of brittle deformation, and, as a result, they overestimate the extent of fluid flow during extension. In contrast to the extensive serpentinization often inferred with bending-related processes during subduction, limited serpentinization is consistent with theoretical predictions and geophysical observations.
\end{abstract}







\end{frontmatter}



\section{Introduction}
Fluid flow in oceanic lithosphere and the resulting chemical alteration are among the important processes occurring at the interface between the Earth’s solid interior and the surface. These processes control the long-term cycling of water and other volatiles, as well as the strength and mechanical behavior of oceanic lithosphere. For example, the water content of subducting oceanic lithosphere, and how much of this water is chemically bound rather than free, influence slab melting, slab seismicity, and how much water is ultimately delivered to the mantle \citep[e.g.,][]{meade1991,green1995,hyndman2003,rupke2004,grove2009}. The delivery of water from subducting oceanic lithosphere to the mantle is a major flux in the deep water cycle \citep{ito1983,jarrard2003}. However, the mechanism controlling the timing and extent of hydration is not obvious because water is buoyant and has to compete with confining pressure, even with plenty of pore space (e.g., cracks and faults). One possibility is that hydration occurs primarily from fluid-filled thermal cracks, which form from thermal stresses during lithospheric cooling and may be tens of kilometers deep \citep{korenaga2007}. With this mechanism, hydrous alteration at shallow depths is expected, but deep serpentinization is less certain and depends on the subtle competition between confining pressure and lithospheric yield strength \citep{korenaga2007,korenaga2017}. In any case, slab hydration would progressively occur as oceanic lithosphere ages and cools. Another possibility concerns the extensive normal faulting that occurs in oceanic lithosphere during subduction \citep[e.g.,][]{christensen1988,ranero2003}. Under this hypothesis, the majority of slab hydration is related to subduction processes, where tectonic stresses transport water down bending-related normal faults, leading to extensive serpentinization \citep[e.g.,][]{ranero2003,faccenda2009,grevemeyer2018}. In this case, mantle hydration only occurs as the slab approaches the trench. Although these two proposed mechanisms are not mutually exclusive, their relative contributions to hydration of oceanic lithosphere are debated, with some arguing that hydration is minimal and occurs primarily at thermal cracks and others that it is extensive and occurs primarily at bending-related normal faults.

Geophysical observations may be employed to investigate when, where, and to what extent hydration occurs, but interpretations are often non-unique. For example, low seismic velocities in oceanic mantle lithosphere have often been interpreted as serpentinization \citep[e.g.,][]{christensen2004,grevemeyer2018}, but low seismic velocities alone cannot distinguish between hydration and crack-like porosity \citep{korenaga2017}. Similarly, the occurrence of intermediate-depth outer rise earthquakes may be explained by dehydration embrittlement following deep hydration \citep{peacock2001}, but both thermal cracking and bending-related normal faulting can potentially facilitate mantle hydration. Also, only a very small amount of serpentinization is required to explain the size-frequency distribution of intermediate-depth earthquakes \citep{korenaga2017}, and it seems possible that either mechanism could facilitate at least this degree of hydration.

Geodynamic modeling has the potential to overcome the uncertainties associated with observations by directly predicting fluid flow in oceanic lithosphere. Modeling results may then be used to infer the extent of chemical alteration. This approach has been taken to estimate fluid flow during plate bending \citep{faccenda2009}, with the result that extensive normal faulting accompanies subduction of an oceanic plate. Darcy’s law is used to track fluid flow along normal faults, e.g., the vertical fluid velocity $v_f$ is given by
\begin{equation}
\label{eq:1}
v_f = -\frac{k}{\phi \eta_f}\left( \frac{\partial P}{\partial z} - g \rho_f \right),
\end{equation}
where $k$ is permeability, $\phi$ is porosity, $\eta_f$ is fluid viscosity, $P$ is pressure, $z$ is depth, $g$ is gravitational acceleration, and $\rho_f$ is fluid density. Here, fluid transport is controlled by the vertical gradient of the difference between full and hydrostatic pressures. This difference is equivalent to the difference between confining pressure ($\rho g z - \rho_f g z$, where $\rho$ is rock density) and dynamic pressure ($P - \rho g z$). In order to achieve downward water transport, the effect of dynamic pressure due to bending must overcome that of the fluid buoyancy. The dynamic pressure resulting from slab bending in the model of \citet{faccenda2009} is indeed sufficient to transport water down faults to mantle depths. This result has been taken as compelling support for the idea that extensive deep hydration is facilitated near the trench by bending-related faulting. However, the numerical model appears to exceed a theoretical limit on the magnitude of dynamic pressure that can be generated in extensional settings such as bending-related normal faulting \citep{korenaga2017}. Because fluid flow is controlled by pressure, this casts doubt on conclusions drawn about extensive mantle hydration. Thus, the relative importance of thermal cracking and bending-related normal faulting in hydrating oceanic lithosphere, and the ultimate extent of mantle hydration, are still open questions considering the current state of geophysical observations and geodynamic modeling.

In the following, we revisit the theoretical arguments of \citet{korenaga2017} in the context of some simple numerical demonstrations in order to examine the ability of geodynamic models to accurately model pressure and, therefore, fluid flow. We then discuss the implications of our demonstration for previous modeling results and provide a recommendation for future modeling studies. We also carefully review relevant geophysical observations of oceanic lithosphere, and finally we provide an outlook on observational and modeling studies concerning the fate of fluids in oceanic lithosphere.

\section{Dynamic pressure in geodynamic models using the pseudoplastic approach}
\label{sec1}
Because fluid flow is governed by the pressure field, it is important to
accurately model pressure when considering fluid flow in geodynamic models.
To evaluate the ability of geodynamic models to solve for pressure accurately,
we generate simple models of horizontal extension and compare the results
to theoretical prediction of dynamic pressure derived from Anderson’s
theory.

\subsection{Numerical method}
We solve the two-dimensional equations for conservation of mass and momentum with incompressibility and viscous flow: 
\begin{equation}
\label{eq:2}
\nabla \cdot \mathbf{u} = \mathbf{0},
\end{equation}
\begin{equation}
\label{eq:3}
- \nabla P + \nabla \cdot \left[ \eta \left( \nabla \mathbf{u} + \nabla \mathbf{u}^T \right) \right] + \rho g \mathbf{e_z} = \mathbf{0},
\end{equation}
where $\mathbf{u}$ is velocity, $P$ is the full pressure, $\eta$ is viscosity, $\rho$ is density, $g$ is gravitational acceleration, and $\mathbf{e_z}$ is the unit vector in the $z$ direction. We test two different solvers, one using the finite element method and one using the finite difference method with a staggered grid. The latter is after \citet{gerya2007} and is used by \citet{faccenda2009}. We only show the results of the finite difference solution, as the two solvers produce similar accuracy in terms of pressure. In order to model brittle deformation, the pseudoplastic approach is often used \citep[e.g.,][]{moresi1998,gerya2007,van2008,korenaga2010,weller2012}, wherein the plastic yield stress is used to define an effective viscosity. This allows for the Stokes equations to capture both viscous and brittle deformation. For example, a ``plastic viscosity'' $\eta_p$ may be defined by equating the yield stress with the actual stress, in this case the second invariant of the deviatoric stress:
\begin{equation}
\label{eq:4}
\eta_p = \frac{\sigma_y}{2 \dot e_{II}},
\end{equation}
where $\sigma_y$ is the yield stress and $\dot e_{II}$ is the second invariant of the strain rate tensor. Here, we define the yield strength using the Coulomb failure criterion:
\begin{equation}
\label{eq:5}
\sigma_y = C + \mu P,
\end{equation}
where $C$ is cohesive strength and $\mu$ is the friction coefficient. Then, $\eta_p$ and the prescribed material viscosity, $\eta_v$, may be combined to give an effective viscosity that is used in eq.~\ref{eq:3}. Specifically, $\eta_p$ should be used wherever the yield stress is exceeded, which is equivalent to the following:
\begin{equation}
\label{eq:6}
\eta = \mathrm{min} \left[\eta_v, \eta_p \right].
\end{equation}
Because the pseudoplastic approach involves a stress-dependent effective viscosity, self-consistency requires iteratively computing the numerical viscosity and solving the Stokes equations until the solution converges. We uniformly prescribe the following material properties: $\rho = 3300$ kg m$^{-3}$, $g = 10$ m s$^2$, $\eta_v = 10^{25}$ Pa s, $C = 0$ Pa, $\mu =  0.6$.

To facilitate horizontal extension, we apply the following velocity boundary conditions:
\begin{equation}
\label{eq:7}
\begin{split}
u_x = -r/2 & \:\:\: \mathrm{on} \:\:\: x=0 \\
u_x = r/2 & \:\:\: \mathrm{on} \:\:\: x=L_x \\
u_z = 0 & \:\:\: \mathrm{on} \:\:\: z=0 \\
u_x = -r L_z / L_x & \:\:\:  \mathrm{on} \:\:\: z=L_z,
\end{split}
\end{equation}
where $r = 2$ cm yr$^{-1}$ is the extension velocity, $L_x = 400$ km is the horizontal domain size, and $L_z = 300$ km is the vertical domain size. The domain is discretized into a grid of $420\times100$ elements. The result of modeling horizontal extension with pseudoplastic rheology is that, over time, linear weak zones emerge. These weak zones are of inherently finite thickness and are therefore an approximation of normal faults. They are characterized by low effective viscosity, strain localization, and alignment with the optimal fault dip. It has previously been pointed out that some particular characteristics of these weak zones, such as their onset time, exact locations, orientation, and thickness, depend on the mesh and numerical method used \citep[e.g.,][]{choi2015,buiter2016,duretz2019}. Thus, although conclusions should not be drawn from such specific characteristics, it is reasonable to assume that modeling results broadly reflect deformational characteristics and may be used to infer the extent of fluid flow.

\subsection{Theoretical prediction and comparison with numerical results}
Here, we derive a theoretical prediction for dynamic pressure due to normal faulting in order to test the assumption that models using the pseudoplastic approach can accurately predict dynamic pressure and, in turn, fluid flow. Two-dimensional plane strain and incompressibility are assumed in the following analysis in order to facilitate comparison with our numerical results.

\begin{figure}[h]
\centering
\includegraphics[width=0.6\textwidth]{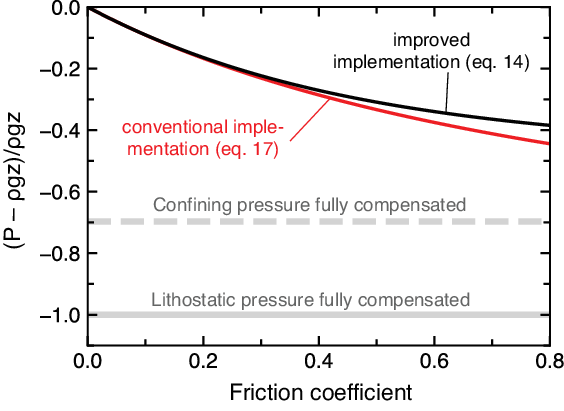}
\caption{Prediction of tectonic pressure (dynamic pressure normalized by the lithostatic pressure) as a function of friction coefficient, according to the assumptions involved in the conventional numerical implementation of pseudoplasticity (red curve) and the improved implementation (black curve). The improved implementation corresponds to the theoretical prediction for pressure due to horizontal extension. Here, the optimal fault dip is assumed. The dashed and solid gray lines correspond to when dynamic pressure compensates confining and lithostatic pressure, respectively. Confining pressure is $\left( \rho - \rho_w \right) g z$, where $\rho_w$ is water density. Tectonic pressure reaches $-1.0$ in the bending models of \citet{faccenda2009} and $-0.6$ in the simple extension models presented here (Fig.~\ref{fig:2}).}\label{fig:1}
\end{figure}

The normal stress components $\sigma_{xx}$ and $\sigma_{zz}$ may be decomposed into volumetric and deviatoric parts:
\begin{equation}
\label{eq:8}
\begin{split}
\sigma_{xx} & = -P + \sigma'_{xx} \\
\sigma_{zz} & = -P + \sigma'_{zz},
\end{split}
\end{equation}
where the primes indicate deviatoric stress. For two-dimensional plane strain, pressure is defined as 
\begin{equation}
\label{eq:9}
P = - \frac{1}{2}\left(\sigma_{xx} + \sigma_{zz} \right).
\end{equation}
For simple horizontal extension, the principal stresses are $\sigma_{xx}$ and $\sigma_{zz}$, and $\sigma_{zz}$ is equal to the lithostatic pressure:
\begin{equation}
\label{eq:10}
\sigma_{zz} = -\rho g z.
\end{equation}
The normal and shear stresses, $\sigma_n$ and $\sigma_s$, on a fault with dip $\beta$ may be expressed in terms of the principal stresses \citep[e.g.,][]{turcotte2002}:
\begin{equation}
\label{eq:11}
\begin{split}
\sigma_n & = \frac{1}{2} \left(\sigma_{xx} + \sigma_{zz} \right) - \frac{1}{2} \left(\sigma_{xx} - \sigma_{zz} \right) \cos 2\beta \\
\sigma_s & = - \frac{1}{2} \left(\sigma_{xx} - \sigma_{zz} \right) \sin 2\beta.
\end{split}
\end{equation}
The condition for frictional failure relates the shear and normal stresses:
\begin{equation}
\label{eq:12}
|\sigma_s| = C + \mu \left( \sigma_n - P_h \right) = \mu \sigma_n,
\end{equation}
where $P_h$ is the hydrostatic pressure. Here, we have assumed $C=0$, as is prescribed in our numerical models, and zero hydrostatic pressure, as fluid is not explicitly included in the models. Equations \ref{eq:8}--\ref{eq:12} may be combined and solved for pressure as a function of lithostatic pressure:
\begin{equation}
\label{eq:13}
P = \frac{\left( \sin 2\beta - \mu \cos 2\beta \right) \rho g z}{\sin 2\beta + \mu \left(1-\cos 2\beta \right)}.
\end{equation}
We may then solve for theoretical predictions of dynamic pressure, defined as $P_D = P – \rho g z$, and tectonic pressure, defined as $P_T = P_D/\left( \rho g z \right)$. Note that tectonic pressure is actually a ratio of pressures and is therefore nondimensional. The theoretical prediction for dynamic pressure due to normal faulting is
\begin{equation}
\label{eq:14}
P_D = \frac{- \mu \rho g z}{\sin 2\beta + \mu \left(1-\cos 2\beta \right)}.
\end{equation}
The more general form of this theoretical prediction was given by \citet{korenaga2017}. Eq.~\ref{eq:14} is relevant for the specific conditions applied in our numerical modeling, such as two-dimensional plane strain. The prediction may be further simplified if the optimal fault dip is assumed. This is the angle at which the deviatoric stress $\sigma'_{xx}$ is minimized, which occurs when $\tan 2\beta = -1/ \mu$ for the extensional case \citep{turcotte2002}. The dependence of dynamic pressure on friction coefficient (according to eq.~\ref{eq:14} and the optimal fault dip) is shown by the black curve in Fig.~\ref{fig:1}. For $\mu=0.6$, dynamic pressure is predicted to be $-0.34\rho g z$. Even if the actual dip angle differs significantly from optimal, the predicted dynamic pressure is similar \citep[see Fig. 1a in][]{korenaga2017}.

\begin{figure}[h]
\centering
\includegraphics[width=1.2\textwidth]{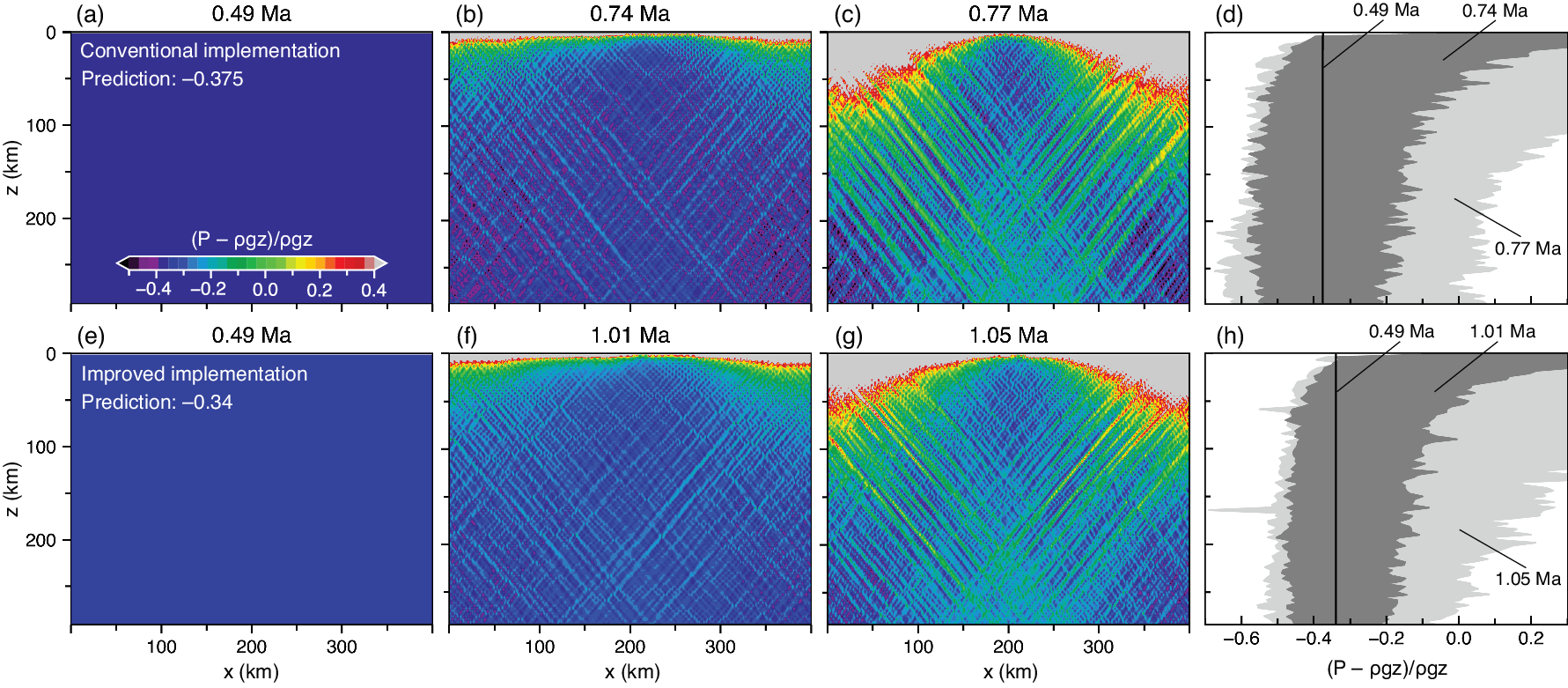}
\caption{Pressure from numerical models of horizontal extension. The top row corresponds to the conventional implementation of pseudoplasticity, and the bottom row corresponds to the improved implementation. Each of the two implementations correspond to a different prediction for pressure (see Fig.~\ref{fig:1}). The first three panels in each row show snapshots of tectonic pressure at different timesteps, and the last panel shows the range of tectonic pressures present at each timestep as a function of depth. Before weak zones form, the pressure field exactly matches the prediction. Note that weak zones form at different times in the two cases, because weak zone nucleation is sensitive to the details of implementation. The increase in tectonic pressure near the surface is due to the velocity boundary condition (see Supplementary Material), and pressures away from the surface are most relevant for comparison to theory.}\label{fig:2}
\end{figure}

This theoretical prediction of dynamic pressure can be used as to benchmark the numerical results corresponding to the conventional approach outlined in the previous section (Fig.~\ref{fig:2}, top row). It is important to note that the increase in dynamic pressure near the surface is due to the surface boundary condition assumed here ($u_z = 0$ on $z=0$). Models incorporating a free surface do not exhibit such high surface pressures (see Supplementary Material) because weak zones can fully extend and accommodate extension near the surface. In any case, once weak zones form, dynamic pressure away from the surface becomes more negative than the value of $-0.34\rho g z$ predicted by eq.~\ref{eq:14}. What’s more, during the period before weak zones form (Fig.~\ref{fig:2}a), the tectonic pressure field is uniform, but even this uniform value of tectonic pressure does not match the theoretical prediction. The reason for this is apparent when the details of the numerical implementation are reconsidered. In theory, the condition for failure is given by eq.~\ref{eq:12}, but the numerical implementation involves equating the yield stress with the second invariant of deviatoric stress, $\sigma'_{II}$:
\begin{equation}
\label{eq:15}
\sigma'_{II} = \mu P.
\end{equation} 
This approach is taken for practicality and simplicity, but it is inconsistent with the assumptions of the above theoretical prediction, and it leads to a discrepancy between the numerical dynamic pressure and the prediction of eq.~\ref{eq:14}. For a more appropriate comparison with the numerical results, we may derive a second ``theoretical'' prediction that is consistent with eq.~\ref{eq:15}. The second invariant of deviatoric stress is given by
\begin{equation}
\label{eq:16}
\sigma'_{II} = \sqrt{\frac{1}{2}{\sigma'_{xx}}^2 + \frac{1}{2}{\sigma'_{zz}}^2 + \sigma_{xz}^2} = \sigma'_{xx}.
\end{equation} 
Two-dimensional incompressibility implies ${\sigma'_{zz}}^2 = {\sigma'_{xx}}^2$, and horizontal extension implies the principal stresses are $\sigma_{xx}$ and $\sigma_{zz}$, and therefore $\sigma_{xz} = 0$. Using eqs.~\ref{eq:15} and \ref{eq:16} along with eqs.~\ref{eq:8}--\ref{eq:10} produces a new theoretical prediction that is more consistent with the numerical implementation:
\begin{equation}
\label{eq:17}
P_D = \frac{-\mu \rho g z}{1+\mu}.
\end{equation} 
Here, $\mu=0.6$ implies $P_D = -0.375\rho g z$. Before the formation of weak zones, the numerical pressure field exactly matches this prediction (Fig.~\ref{fig:2}a,d), confirming the relevance of eq.~\ref{eq:17} to the conventional numerical implementation.

\begin{figure}[h]
\centering
\includegraphics[width=1.2\textwidth]{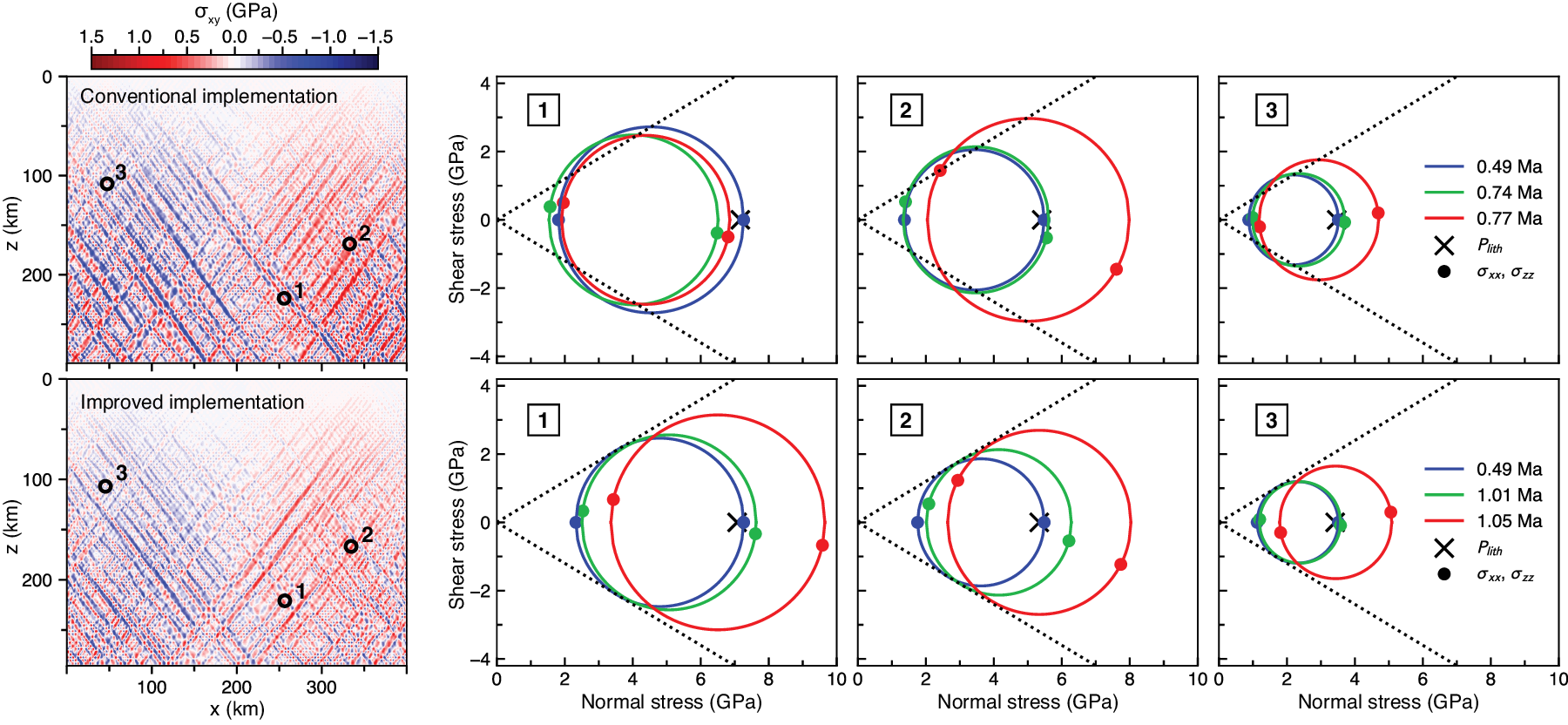}
\caption{Stress from numerical models of horizontal extension. The top row corresponds to the conventional implementation of pseudoplasticity, and the bottom row corresponds to the improved implementation. The first panel in each row shows a snapshot of the shear stress, $\sigma_{xy}$. Large $\sigma_{xy}$ indicates that the principal stresses are not $\sigma_{xx}$ and $\sigma_{zz}$. Numbered circles indicate the locations for which the full stress state is shown in the last three panels of each row. For each of the three locations, we plot stress circles for three different timesteps. To calculate shear and normal stresses, we determine the principal stresses (which may be different from $\sigma_{xx}$ and $\sigma_{zz}$) from the stress tensor. The lithostatic stress is shown with a black `x', and $\sigma_{xx}$ and $\sigma_{zz}$ are shown with filled circles. The dotted lines correspond to the assumed friction law.}\label{fig:3}
\end{figure}

As soon as weak zones form, however, it is apparent that the magnitude of the predicted dynamic pressure is exceeded (Fig.~\ref{fig:2}b-d). The reason for this is not immediately obvious, but further insight is provided by looking at the state of stress at specific locations within weak zones (Fig.~\ref{fig:3}, top row). Constructing stress circles shows that the stress state is not actually bounded by the frictional failure condition. This suggests that the numerical approach of applying the yield stress to the second invariant of deviatoric stress is not consistent with Mohr-Coulomb failure. Instead of modifying the theoretical prediction to match this approach, we are better off modifying the numerical approach to match the original theoretical prediction, which is based on our theories of stress and frictional failure. We can do so by directly applying eqs.~\ref{eq:8}--\ref{eq:12} in the following manner. Those equations may be solved for $\sigma'_{xx}$ in terms of the full pressure: 
\begin{equation}
\label{eq:18}
\sigma'_{xx} = \frac{\mu P}{\left(1+\mu^2 \right)^{1/2}}.
\end{equation} 
This can be interpreted as a form of the yield criterion: $\sigma'_{xx}$ may not exceed eq.~\ref{eq:18}. Thus, we can use this to define an effective viscosity by considering this single component of stress:
\begin{equation}
\label{eq:19}
\eta_p = \frac{\sigma'_{xx}}{2 \dot e_{xx}},
\end{equation} 
where $\sigma'_{xx}$ is given by eq.~\ref{eq:18} and the numerical pressure and strain rate are used to determine $\eta_p$.

Using this new implementation, the state of stress at yielding is consistent with the failure criterion (Fig.~\ref{fig:3}, bottom row). This demonstrates that numerical implementation is important for ensuring the failure criterion is enforced, and that the conventional implementation of plastic yielding, which uses the second invariant of stress, is not sufficient. However, when the dynamic pressure field is considered, the theoretical prediction is still exceeded (Fig.~\ref{fig:2}f-h). This is puzzling because the failure criterion is exactly satisfied in the weak zones and the theoretical prediction of dynamic pressure is based on the assumption of this criterion. Looking at the stress state within weak zones over time (Fig.~\ref{fig:3}, bottom row) reveals that, after weak zones form, the principal stresses can rotate significantly from $\sigma_{xx}$ and $\sigma_{zz}$ (indicated by large $\sigma_{xz}$), and $\sigma_{zz}$ can be far from the lithostatic stress. However, theory predicts that, for horizontal extension, principal stresses are $\sigma_{xx}$ and $\sigma_{zz}$, and $\sigma_{zz} = -\rho g z$, which leads to the dynamic pressure given by eq.~\ref{eq:14}. It seems that, because the weak zones have a finite thickness and are modeled as viscously deforming material, the principal stresses rotate away from the horizontal and vertical axes. Thus, the culprit is the pseudoplastic approximation itself, which models faults (i.e., discontinuities) originating from brittle deformation as finite-thickness weak zones originating from viscous deformation. \citet{mancktelow2008} observed that rotation of principal stresses is common in layers of viscously deforming material with contrasting viscosities. This applies to the weak zones resulting from the pseudoplastic approach, which are layers with low effective viscosity bounded by high-viscosity material. Principal stress rotation has also been predicted in a theory of plastic strain localization \citep{lepourhiet2013}. We therefore argue that the pseudoplastic approach, the most common approach for representing brittle deformation in geodynamic models, is insufficient for predicting some aspects of brittle deformation, including the pressure field, which is used to infer fluid flow.

\section{Implications for modeling fluid flow}
We have demonstrated that dynamic pressure due to brittle deformation is not appropriately modeled using the pseudoplastic approach. This motivates reconsideration of previous modeling based on this approach, particularly those that model fluid flow from the numerical pressure. The numerical implementation used by \citet{faccenda2009,faccenda2012} involves the pseudoplastic approach in the sense that brittle deformation is modeled by using an effective viscosity in the Stokes equations \citep{gerya2007}. In their numerical bending simulations, tectonic pressure of up to $-100 \% $ (i.e., dynamic pressure equal to the lithostatic pressure) is reported and can be seen in their Fig.~\ref{fig:2}, which is more than double the theoretical prediction for normal faulting with a friction coefficient of 0.6 (Fig.~\ref{fig:1}). This numerical pressure is then used to predict fluid flow using eq.~\ref{eq:1}, with the result that large negative tectonic pressures enhance the downward transport of fluids along normal faults. However, the numerical demonstration in the previous section implies that this large tectonic underpressure—and in turn, the extensive fluid flow—is likely overestimated.

We offer the following recommendations for representing brittle deformation in geodynamic models. First, the pseudoplastic approach is still useful for investigating the qualitative characteristics of brittle deformation. For example, the pseudoplastic approach is likely sufficient for implementing the large-scale brittle weakening of oceanic lithosphere required for achieving plate-tectonic convection in geodynamic models. When using the pseudoplastic approach, it is best to choose an implementation consistent with the chosen failure criterion. For example, an appropriate implementation of the frictional failure criterion is given in the previous section (see eq.~\ref{eq:19}). However, it is important not to interpret the magnitude of pressure resulting from pseudoplasticity, as it appears that this issue is inherent to modeling faults as shear bands. For example, \citet{duretz2020} model shear bands during horizontal extension using a new implementation of plasticity. When their numerical pressure is converted to tectonic pressure, the theoretical prediction is exceeded by up to 30\%, which is similar to the discrepancy seen in the models presented here with conventional pseudoplasticity. This supports the idea that a discrepancy with theoretical pressure is inherent to modeling frictional failure with plastic shear bands. We recommend that, if pressure magnitudes are important to the modeling result, a theoretical prediction for pressure should be applied as a bound on the numerical pressure. Alternatively, modeling faults explicitly (e.g., with mesh discontinuities as is common in the fracture mechanics community) will avoid this issue to begin with, at the cost of increased numerical complexity.

\section{Geophysical observations linked to fluid flow in oceanic lithosphere}
The previous section demonstrates that geodynamic models have overestimated the extent of fluid flow due to bending-related faulting. As these models are often invoked as the geodynamic basis for the interpretation of geophysical observations, a reconsideration of observational arguments is warranted and may clarify what is possible with respect to the extent and nature of deep serpentinization.

\subsection{Anomalous seismic velocities}
Bending of subducting oceanic lithosphere inevitably introduces extensive systems of normal faulting, and such faulting was first seismically imaged a few decades ago \citep{ranero2003}. Estimating the extent of hydration introduced by such bending-related faulting has been a goal of many studies in active-source seismology \citep[e.g.,][]{ranero2004,grevemeyer2007,contreras2008,ivandic2008,van2011,contreras2011,lefeldt2012,fujie2013,shillington2015,grevemeyer2018,fujie2018,wan2019,he2023,he2025}, and some in passive-source seismology \citep[e.g.,][]{cai2018,zhu2021,mark2023,li2024}. Models using active-source data are able to constrain seismic velocities of the oceanic crust and, at most, the top $\sim$10 km of the mantle. Because of poor S-wave data quality, most of these studies consider only P-wave velocities. The S-wave models that have been published are either limited to the crust or include only the top few kilometers of mantle. Passive-source studies typically involve surface wave tomography and thus only generate S-wave models, although P-wave velocity has been constrained by receiver functions in one case \citep{li2024}. These models can penetrate deeper into the mantle, but are more limited in resolution than those constrained from active-source data. Active- and passive-source studies consistently find reduced seismic velocities in at least the top few kilometers of oceanic mantle, with a typical P-wave velocity reduction of $10\%$. The most popular interpretation of these observations is serpentinization, and the idea that bending-related processes extensively serpentinize oceanic mantle has gained significant acceptance \citep[e.g.,][]{morgan2023}. However, our numerical demonstration casts doubt on the physical basis of this idea, motivating a reconsideration of the seismic observations.
\begin{figure}[h]
\centering
\includegraphics[width=1.0\textwidth]{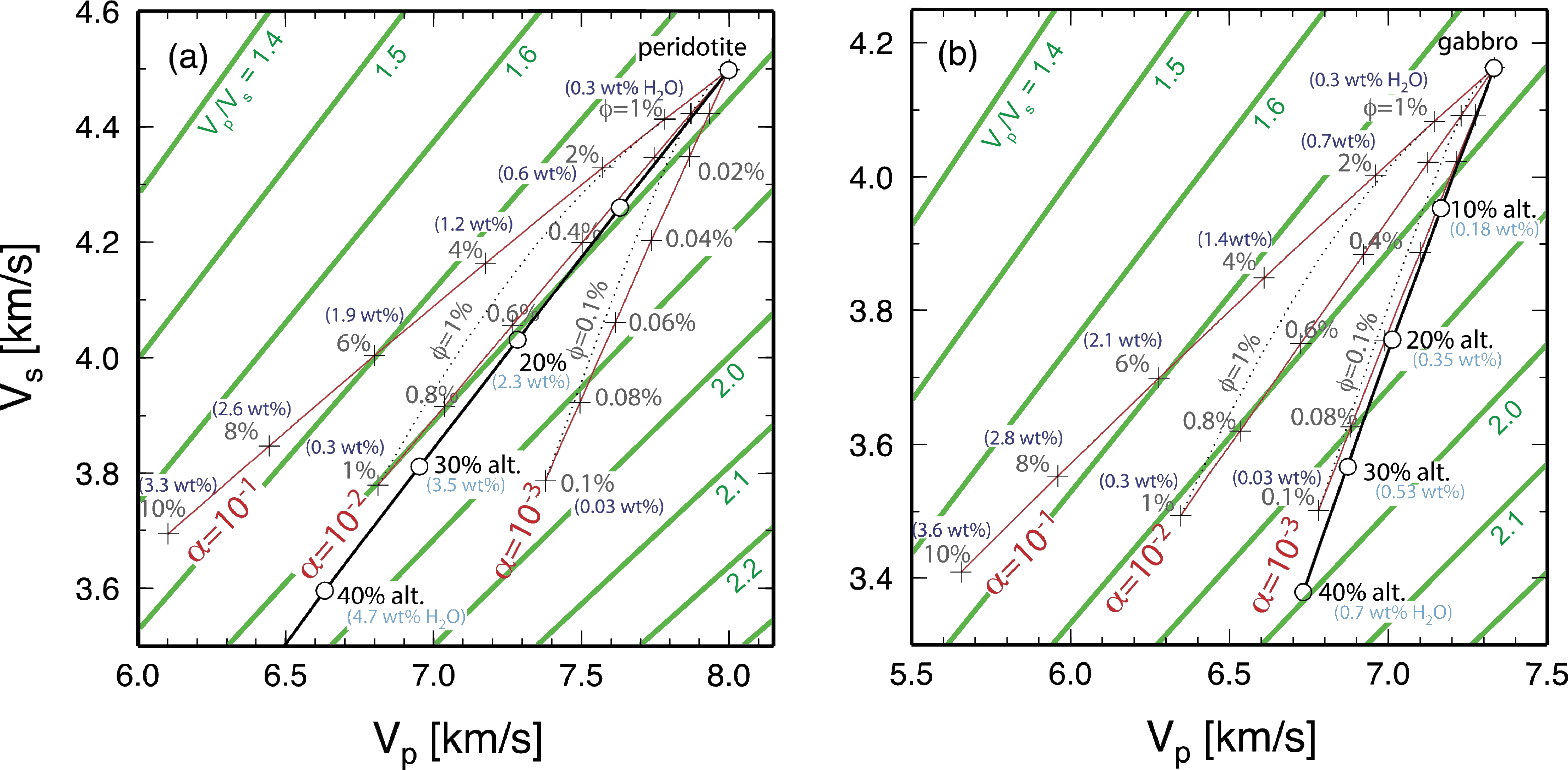}
\caption{Effects of chemical alteration (black lines) and crack-like porosity (red lines) on seismic velocities of peridotite (left) and gabbro (right), after \citet{korenaga2017}. Numbers (sans parentheses) indicate either the degree of alteration or porosity, and numbers in parentheses indicate the corresponding water content. Green contours indicate Vp/Vs at intervals of 0.1. See \citet{korenaga2017} for the details of the calculations.}\label{fig:4}
\end{figure}

Extensive serpentinization ($\sim$20--30$\%$) is required to explain a P-wave velocity reduction of $10\%$ (Fig.~\ref{fig:4}a), corresponding to $\sim$2--3 wt$\%$ water. The same velocity reduction can be explained by just a small amount of crack-like porosity, with the exact amount of required porosity depending on the crack aspect ratio (Fig.~\ref{fig:4}a). For example, if aspect ratio is 10$^{-2}$, less than 1$\%$ porosity is required to explain a $10\%$ velocity reduction. Aspect ratio refers to the ratio between the short and long semi-axes, so a smaller aspect ratio means more crack-like geometry. Some of the studies that include S-wave models have aimed to eliminate the nonuniqueness of interpretations using constraints on Vp/Vs \citep{grevemeyer2018,he2025}. For example, \citet{grevemeyer2018} observe increasing Vp/Vs with decreasing velocities and argue that only serpentinization can explain this trend, because Vp/Vs is expected to increase with serpentinization while it is expected to decrease with porosity. However, \citet{korenaga2017} pointed out that the trend in Vp/Vs due to porosity depends on the crack aspect ratio. For example, if aspect ratio is 0.1, Vp/Vs decreases with porosity, but if aspect ratio is 10$^{-2}$ or smaller, Vp/Vs increases with porosity (Fig.~\ref{fig:4}a). Importantly, the velocities expected for serpentinization are only consistent with a small range of Vp/Vs, with 20--30$\%$ serpentinization corresponding to Vp/Vs of $\sim$1.80--1.82 (Fig.~\ref{fig:4}a). Thus, Vp/Vs significantly different from 1.8 along with 10$\%$ velocity reduction could only be explained by crack-like porosity. Notably, \citet{li2024} report Vp/Vs greater than 2.05 in the mantle of the incoming plate at the Alaska subduction zone. In this case, seismic properties are uniquely explained by crack-like porosity. The values of Vp/Vs observed by \citet{grevemeyer2018} and \citet{he2025}, on the other hand, are consistent with either serpentinization or crack-like porosity with aspect ratio of $\sim$10$^{-2}$. The two interpretations correspond to considerably different water content. For example, 30$\%$ serpentinization or 1$\%$ porosity can explain the same seismic properties, but the former corresponds to $\sim$3 wt$\%$ water and the latter corresponds to $\sim$0.3 wt$\%$ (Fig.~\ref{fig:4}a). More and better S-wave models are needed in order to constrain a unique interpretation of the reduced seismic velocities in the subducting oceanic mantle.

In addition to the oceanic mantle lithosphere, the lower oceanic crust also frequently exhibits reduced velocities and/or elevated Vp/Vs \citep[e.g.,][]{ranero2004,van2011,fujie2013,fujie2018,li2024,he2025}. Even before entering subduction zones, normal oceanic lower crust has P-wave velocities of $\sim$7.0 km/s \citep{white1992}, which is low compared to the $\sim$7.3 km/s expected for unaltered, unfractured gabbro (Fig.~\ref{fig:4}b). Both alteration \citep{carlson2003} and crack-like porosity \citep{korenaga2002} have been suggested to explain this discrepancy. Either way, only P-wave velocities lower than $\sim$7.0 km/s should be interpreted as anomalous with respect to normal lower crust. Where this is observed, it is tempting to infer pervasive bending-related hydration. This is because, for gabbro, the velocity reduction introduced by alteration is only moderate, so that even a small velocity reduction could imply a large degree of alteration (Fig.~\ref{fig:4}b). However, because the olivine content of gabbro is small, the amount of water introduced by such alteration is nearly an order of magnitude smaller for gabbro than for peridotite. For example, 30$\%$ alteration of gabbro corresponds to only $\sim$0.5 wt$\%$ water, compared to $\sim$3.5 wt$\%$ for peridotite. Thus, the crust has a limited capacity for storing chemically-bound water. Further, if fluid-filled porosity, instead of alteration, is responsible for some or all of the velocity decrease, then water content will be much smaller than for alteration alone. For example, a 6$\%$ decrease in Vp could be explained by 30$\%$ alteration or just 0.1$\%$ porosity (with aspect ratio of 10$^{-3}$), which correspond to 0.5 and 0.03 wt$\%$ water, respectively (Fig.~\ref{fig:4}b). Regarding Vp/Vs of the lower crust, a few studies report values greater than $\sim$1.85 \citep[e.g.,][]{fujie2018,he2025}, which is higher expected for unaltered, unfractured gabbro (Fig.~\ref{fig:4}b). The high Vp/Vs has been attributed to pervasive fluid flow and chemical alteration. However, because the amount of crack-like porosity required to explain Vp/Vs $>$ 1.85 is very small (0.1$\%$ or less), bending faults alone, i.e., without alteration, can explain the seismic observations. Some alteration cannot be ruled out and is probably expected, but the water content introduced by such alteration is quite limited, as noted above.

In order to infer degree of hydration in the incoming plate from seismic velocities, bulk serpentinization is often assumed, either implicitly or explicitly. However, the extent of serpentinization due to fracturing is controlled by the ability of water to flow horizontally through wall rock, which in turn is controlled by microporosity of wall rock and is likely limited at large lithostatic pressures \citep[see section 3.2 of][]{korenaga2017}. \citet{hatakeyama2017} measured the permeability of serpentinites and estimated that the resulting lateral extent of serpentinization is on the order of hundreds of meters. However, their calculation assumes that fluid flow is controlled by equilibrium vapor pressure at the reaction front, but it is more likely to be controlled by the confining pressure \citep[see section 3.2 in][]{korenaga2017}, which quickly increases with depth. Thus, where serpentinization does occur, it is likely closely confined to the faults themselves. This limited extent of serpentinization can more easily explain seismic observations than bulk serpentinization \citep{miller2016}. That is, an order of magnitude lower water content is implied by serpentinization confined to fault zones, as opposed to bulk serpentinization. Thus, the degree of hydration suggested by many observational studies, which can exceed 30$\%$ serpentinization and 3.5 wt$\%$ water, are not only upper bounds but likely considerable overestimates. Observations of seismic anisotropy can potentially constrain the extent of serpentinization around fault zones. For example, observations by \citet{miller2021} of azimuthal anisotropy in the outer rise upper mantle are incompatible with uniform serpentinization, but can be explained by serpentinization limited to zones less than 100 m thick. Fractured, unaltered rock can also introduce anisotropy, but differences in seismic signals expected for thin serpentinized fault zones versus fracturing alone are subtle \citep{miller2021}. Thus, we emphasize the importance of obtaining more and better constraints on seismic anisotropy in the outer rise upper mantle.

\subsection{Intermediate-depth seismicity}
Intermediate-depth earthquakes are those that occur at depths of $\sim$50--200 km in subduction zones. These events generally occur in an upper and lower plane within the bending plate, forming the double seismic zone. The origin of intermediate-depth earthquakes is not immediately obvious, because pressure and temperatures at such depths should prevent frictional sliding. Several decades ago it was proposed that dehydration embrittlement, involving the breakdown of hydrous minerals at depth, is responsible for intermediate-depth seismicity \citep{seno1996,peacock2001,yamasaki2003}, and this is currently the most commonly invoked mechanism to explain intermediate-depth seismicity. Dehydration embrittlement requires that the incoming plate be hydrated to at least some degree. The upper-plane earthquakes, which occur at the top of the plate, are easiest to explain, because this is where hydrous minerals are expected to be most concentrated, due to hydrothermal alteration at mid-ocean ridges or subsequent faulting \citep{kirby1996,hacker2003}. Lower-plane earthquakes from dehydration embrittlement are a consequence of hydration of the upper mantle. Although this may seem to be at odds with the arguments for minimal mantle hydration put forth in the previous section, \citet{korenaga2017} calculated that only an exceedingly small amount---as low as $\sim$1.0 wt. ppm---of bound water is required to explain the frequency of lower-plane earthquakes. In the following, we discuss what the characteristics of lower-plane intermediate-depth seismicity may reveal about the mechanism controlling upper mantle hydration.

If bending-related normal faulting induces mantle hydration, subsequent intermediate-depth earthquakes resulting from dehydration reactions are expected to occur in a relatively dense and uniform pattern, since bending fault spacing is on the order of a few kilometers \citep{ranero2003}. However, although the distribution of upper-plane events is relatively dense, lower-plane events have a distinct clustered distribution \citep[e.g.,][]{igarashi2001,kita2010,sippl2018}, even in the most seismically active and densely observed subduction zones such as Japan. This has been interpreted as evidence against an origin of bending-related hydration for lower-plane seismicity \citep{yamasaki2003}. \citet{shillington2015} proposed that interactions between pre-existing plate fabric and bending-related faulting can explain the distribution of lower-plane earthquakes in the Alaska subduction zone, but \citet{korenaga2017} pointed out that this mechanism should also influence the distribution of upper-plane earthquakes, which are independent of plate fabric. The characteristics of thermal cracks may naturally explain the patterns of intermediate-depth seismicity. Thermal cracks are expected to follow a cascade pattern, where shallow cracks are closely spaced and deep cracks are widely spaced, with the deepest cracks ($\sim$50 km deep) spaced $\sim$50 km apart \citep{korenaga2007}. This cascade pattern is expected because deep cracks do not release thermal stress efficiently at shallow depths, so additional shallow cracks are needed to fully release the thermal stresses generated from lithospheric cooling. This expected crack pattern can naturally account for the dense distribution of upper-plane events and the clustered distribution of lower-plane events. According to this explanation, the lower-plane events cluster around the deepest and most widely spaced cracks. If thermal cracks take on a hexagonal planform, then earthquakes due to dehydration embrittlement may occur in a smeared hexagonal pattern \citep{korenaga2017}. However, three-dimensional modeling of thermal cracks is still needed to fully predict the planform of thermal cracking and resulting predictions for intermediate-depth seismicity.

\section{Concluding remarks}
Geodynamic models of oceanic lithosphere have likely overestimated the extent of fluid flow during tectonic processes. As a result, there is no geodynamic basis for a link between bending-related normal faulting and extensive mantle hydration. Future models incorporating fluid flow should implement brittle deformation explicitly, as opposed to using the pseudoplastic approach, and should rigorously benchmark numerical pressure.

Observations support minimal mantle hydration as much as, or in some cases even more than, extensive hydration. Future observational studies should focus on constraining anisotropy and S-wave velocities in the outer rise upper mantle, both of which can potentially resolve the ambiguity between alteration and porosity. Also, the pattern of intermediate-depth seismicity may reflect the mechanism facilitating mantle hydration. Making use of this connection will require better model constraints on the distribution and geometry of faults and cracks corresponding to potential mechanisms.

\section*{Acknowledgements}
We are grateful to two anonymous reviewers for constructive and insightful feedback.

\bibliographystyle{elsarticle-harv} 
\bibliography{references.bib}






\end{document}